\documentclass[twocolumn]{aastex631}

\usepackage{graphicx}

\usepackage{mathptmx}    
\usepackage{latexsym,amsmath,amssymb,natbib}
\usepackage{graphicx}
\usepackage{dcolumn}
\usepackage{bm}
\usepackage{hyperref}
\usepackage{ulem}
\usepackage{txfonts}
\usepackage{textcomp}

\begin{document}

\title{Aggregation of Sub-mm Particles in Strong Electric Fields under Microgravity Conditions}


\author{Felix Jungmann}
\author{Maximilian Kruss}
\author{Jens Teiser}
\author{Gerhard Wurm}

\address{University of Duisburg-Essen, Faculty of Physics, Lotharstr. 1-21,  47057 Duisburg \\ felix.jungmann@uni-due.de}

\begin{abstract}
  Dust emission mechanisms as one aspect of wind-driven particle motion on planetary surfaces are still poorly understood. The microphysics is important though as it determines dust sizes and morphologies which set sedimentation speeds and optical properties. We consider the effects of tribocharging in this context as grains in wind driven granular matter charge significantly. This leads to large electric fields above the granular bed. Airborne dielectric grains are polarized in these electric fields, which leads to attractive forces between grains. To simulate aggregation under these conditions we carried out drop tower experiments using tracer particles, mimicking the gas coupling behavior of small dust grains in terms of high surface to mass ratios and efficient gas drag. Under microgravity, the particles are released into an observation chamber in which an alternating electric field up to 80\,kV/m is applied. Without electric field no aggregation can be observed on timescales of seconds. However, polarization instantly leads to aggregation of particles when the field is switched on and long chains aligned to the electric field form. Scaled to dust entrained into planetary atmospheres, fine and coarse grain fractions might readily form aggregates after being liberated. 
  Under certain natural conditions, aggregates might therefore start chain-like or at least a chain-like appearance is favored. If atmospheric influences on their stability are small, aerodynamic and optical properties might depend on this.

\end{abstract}

\section{Introduction}
Wind-driven particle flows have been subject to a large number of studies working out details of threshold conditions for lifting and quantifying particle fluxes \citep{Bagnold1941, Greeley1985, Merrison2008, Kok2012, Rasmussen2015, Musiolik2018, Kruss2020a, Swann2020}. Models for the threshold friction velocity are essentially balances between gas drag and friction force, the latter being proportional to gravity and particle's surface energy \citet{Shao2000}.
Once in motion, the particle flow can be sustained even at lower wind velocities by saltation as airborne grains impact the soil and liberate other grains \citep{Bagnold1941}. 

However, it is well known that grains charge during collisions \citep{Lacks2011, Waitukaitis2014, MendezHarper2021}. Even on Mars this may be possible \citep{MendezHarper2020}. Wind-driven flows are no exceptions \citep{Merrison2012}. As consequence of these charging processes, strong electric fields of tens of kV/m are generated \citep{Schmidt1998, Zheng2013, Zhang2020}. This applies to saltating flows \citep{Kok2008, Rasmussen2009, Esposito2016} as well as dust devils \citep{Renno2004, Harrison2016, Franzese2018}. Electrification strongly changes the microphysics of the flow in several ways. On the one side, Coulomb attraction or repulsion within the electric field will alter the conditions under which grains are lifted as electrostatic forces are easily dominating over gravity. It will also alter the trajectories on which grains reimpact the soil \citep{Kok2008}.

On the other side, grains within the soil might be prone to electrostatic aggregation \citep{Steinpilz2020a}. These aggregates might be lifted more easily \citep{Merrison2007} or might be hard to lift if the aggregates were too compact or too large to be lifted. 

While saltation of large grains on the 100 micron scale is restricted to some layer above the surface, it also produces the emission of sub-micron range dust particles. Dust is rather adhesive and a detachment of single grains from the ground is therefore not straightforward. There are several possible mechanisms for liberating the dust, ranging from impacting saltating grains over direct entrainment in turbulent eddies to airborne disintegrating aggregates \citep{Renno2008, Klose2013, Klose2017, Musiolik2017, Bila2020}. Quite generally, dust emission is important for global particle transport. Grains from sub-micrometer to a few tens of micrometer play a role here and size distributions vary strongly \citep{Mahowald2014}. On Mars the thin atmosphere limits the size range of suspended particles from typically 1 \textmu m \citep{Kahre2017} to few micrometers in extreme events as global dust storms \citep{Lemmon2019}. Once lifted though, the fraction of small dust that will be suspended also depends on the interactions within the electric field and within the dense particle cloud above the ground. Thus, colliding dust grains might reaggregate again. On Earth, these effects are also discussed to be part of the aggregation of fine volcanic ash \citep{James2002, Telling2012, Telling2013, Cimarelli2014, Mueller2017}.

In a cloud of charged grains, the net charge on each grain is often considered as primary parameter to decide if grains attract or repel each other.
This neglects the role of dipole interaction or dielectrophoresis. For example, one strongly charged grain can generate an electric field, which is sufficient to induce attractive forces between charged grains, even if their charges have the same sign \citep{Feng2000, Matias2018, Steinpilz2020a, Steinpilz2020b}. In this case, induced dipoles provide the binding forces. Extending this idea, particles in the strong electric field above a wind-driven soil are always polarized. These induced dipoles will attract each other. If this attraction leads to efficient aggregation as grains are lofted, the size and morphology of airborne dust will change. In principle, this is in analogy to aggregation of ferromagnetic particles in a magnetic field \citep{Kruss2018, Kruss2020b}.

It is difficult though to study the microphysics in a natural flow, i.e. determining microscopic aggregate morphologies of airborne grains or aggregation itself while the particles move along the wind at the same time and while measuring but not disturbing electric fields. There are some laboratory measurements on the influence of electric fields on particle transport, e.g. by \citet{Rasmussen2009}, but wind tunnel experiments with charged grains are still rare and not without limitations.

Here, we simulate a part of these processes under well constrained laboratory conditions in a microgravity setting. We focus on the aggregation of dielectric grains in a strong electric field comparable to measured fields above ground during sand transport on earth \citep{Schmidt1998}.

\section{Experiment}

\begin{figure}
 \centering 
\includegraphics[width=0.8\columnwidth]{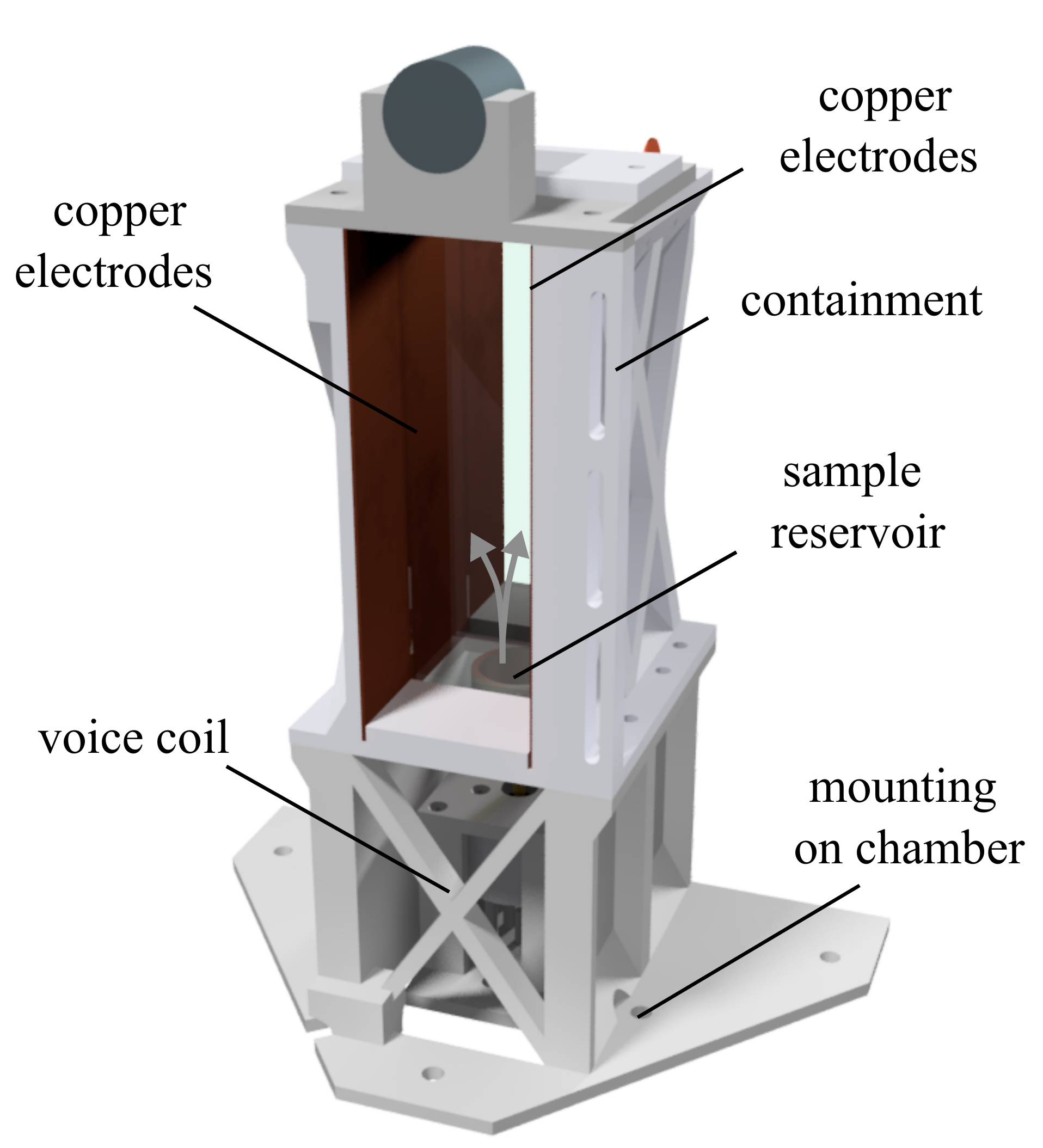}
\caption{Sketch of the main setup of the experiment. Before launch, gravity is pointing downwards so the sample remains in the reservoir. The camera's perspective is rotated by $90^\circ$ around the horizontal axis, therefore, in Fig.~\ref{fig.chain-process} the electrodes are not placed left and right but at the top and the bottom. The grey arrows show the movement of the particles after launch from the reservoir into the capacitor volume}.
\label{fig.setup}
\end{figure}

The basic experimental setup (Fig.~\ref{fig.setup}) has been used in numerous previous works by now \citep{Jungmann2018, Steinpilz2020a, Steinpilz2020b, Jungmann&Wurm2021} with the main difference that, dry air is used here as surrounding gas instead of $\rm{CO}_2$. The setup consists of a shaker driven by a voice coil. The particle volume of the shaker has a diameter of 25\,mm. Prior to each experiment the sample is shaken at 35 Hz, so collisions of the sample grains among each other lead to triboelectric charging. The inside of the shaker volume is coated with the same particles as the sample to guarantee same material contacts and overall neutrality (by first order) within the sample. The sample is shaken for 15\,min before each experimental run, which is sufficient to reach an equilibrium net charge distribution on the grains according to \citet{Wurm2019} and \citet{Steinpilz2020a}. Those previous experiments also showed that the charge distribution of the sample does not significantly change within time scales of hours after shaking.

Experiments are carried out under microgravity at the drop tower in Bremen which provides a free fall time of about 4.7 s for a single drop \citep{VonKampen2006}. By launching the setup upwards with a catapult the microgravity time is essentially doubled to about 9\,s, which is the case here. To investigate the interaction between charged grains in detail a certain minimum size of the grains is necessary, as a large ensemble of particles requires a certain field of view, which automatically limits the optical resolution, as microscopes cannot be used. Thus, microgravity is required to achieve an experiment duration long enough. Gravity would add too large disturbances in view of small Coulomb forces, as settling and convection would dominate the system. The setup is launched after charging the grains before on the ground as described above.
Grains are released into the observation chamber of the experiment after onset of microgravity. This is achieved by agitation of the voice coil at a low frequency  with a large amplitude. The observation chamber measures $50\,x\,50\,x\,120\,\mathrm{mm}^3$ and is at atmospheric pressure during all experiments. The side walls of the chamber act as capacitor. High voltage can be applied to create a homogeneous electric field in the chamber and particles react depending on their charges and dipoles in this field. Applied fields are sinusoidal with 80\,kV/m peak-to-peak and a frequency of 100\,Hz. The particles are illuminated from the back using a diffusor, i.e. using bright field imaging and their motion and aggregation is observed with a camera at the front at a frame rate of 100\,frames/s. The camera's perspective is rotated by $90^\circ$ around the horizontal axis, therefore, in our images, the electrodes are at the top and bottom (Fig.\ref{fig.chain-process}).

The particles used in this study are hollow glass spheres of 150-180\,\textmu m diameter with an average density of 0.072 g/cm$^3$. We used these particles as aerodynamic analogue for real, compact micrometer dust grains. Despite the difference in size, both have comparable gas-grain coupling times, with the hollow spheres easily resolved by the camera in contrast to small dust grains.

In total, 8 experiments have been carried out turning on and off the electric field with different frequencies at different times. While the formation of chains has been observed in each of these experiments, the analysis in the results section is based on two launches. In the first case, the field is turned on after half of the flight-time and in the second case, the field is non-alternating and switched on for the total flight-time.

\section{Results}

\subsection{Chain formation}

\begin{figure}
    \center
	\includegraphics[width=\columnwidth]{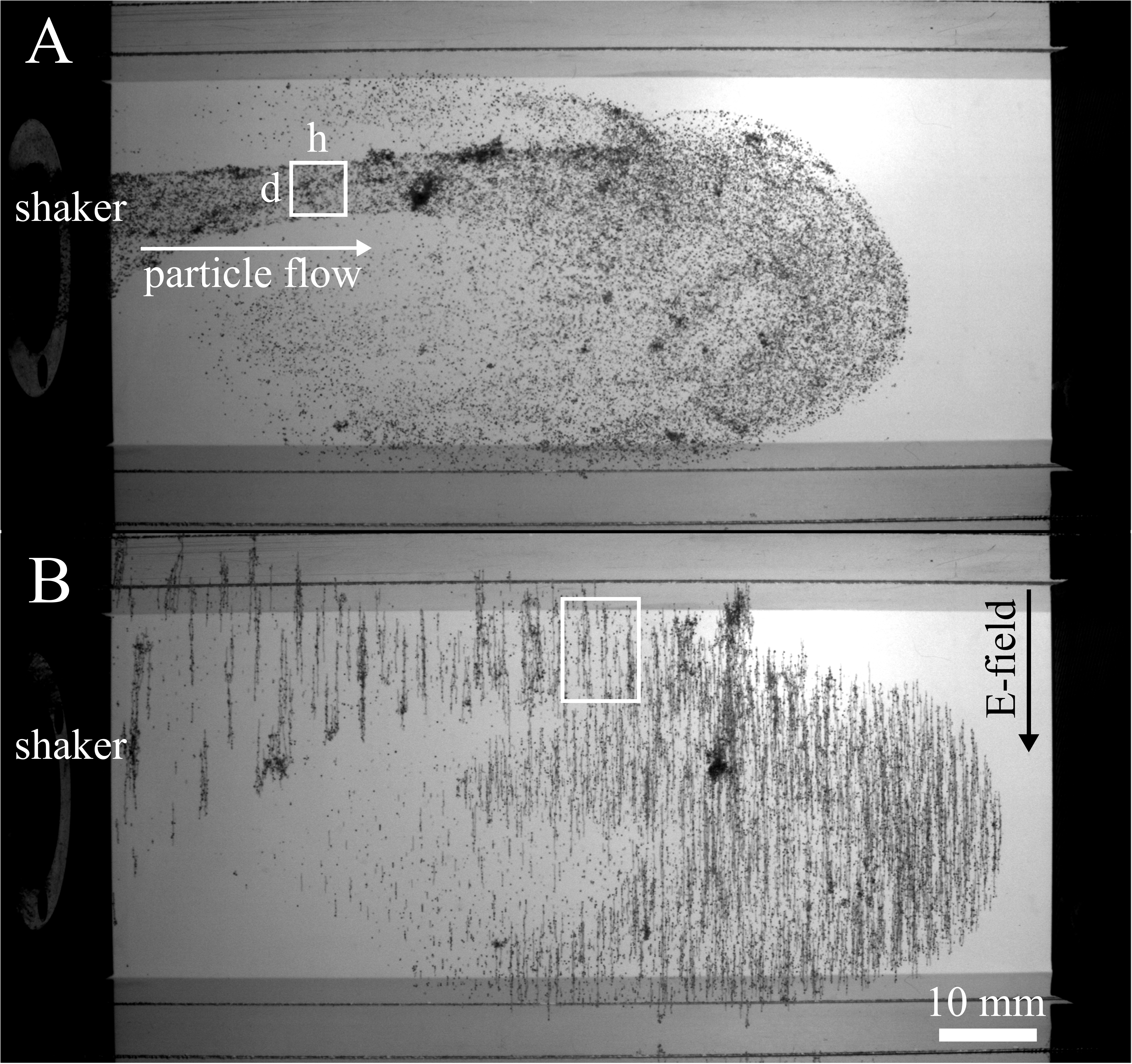}
	\caption{\label{fig.chain-process} Hollow spheres emerging from the shaker. Each image has a resolution of 2240 x 1050 px. The images are tilted by $90^\circ$ with respect to the real setup, so the capacitor plates are top and bottom of the camera images. A: Particle cloud 3.5\,s after the start of microgravity without electric field. The white area marks a region used for density estimates moving along the cloud (see section \ref{chp.agg-model}). B: Particle cloud 3\,s after switch-on of an alternating field of 80\,kV/m and 100\,Hz. The marked region shows the same particles as above. Due to the motion of the cloud and aggregation, its place and dimensions changed. }
\end{figure}

The electric field is initially turned off as the particle cloud emerges into the observation chamber. Some particles already stick to each other due to cohesion and close contacts in the sample reservoir as seen in Fig.~\ref{fig.chain-process}. In this case agglomerates can be distinguished from observational artefacts, as we trace the movement of the agglomerates. In case of an observational artefact (particle occlusion due to 2D-observation), a relative motion between grains will be visible. No sign of additional active clustering due to charge can be observed.
The cloud motion is quickly damped as hollow spheres couple to the gas rapidly and grains then just follow the residual gas motion (Fig.~\ref{fig.chain-process} top). This gas motion is caused by the large amplitude of the voice coil's oscillation (up to 10 mm) and the initial drag by the emerging particle cloud. The particles are now essentially at rest with respect to each other. Net charges on the glass spheres might lead to attraction or repulsion but without external electric field no significant relative motion of neighboring particles can be observed on a timescale of seconds. Consequently, the cloud morphology does not change.

However, once electric fields are applied, chains of particles form instantly on timescales of ms (Fig.~\ref{fig.chain-process} bottom). All chains are aligned parallel to the electric field and perpendicular to the initial motion of the particles. The fast alternating electric field prevents charged grains from accelerating towards one of the electrodes.
We note that this is different from  real dust events on a planetary surface, where E-fields only change on longer timescales. Therefore, in reality, net charges will contribute as well to the overall motion in the direction of the E-field. At high frequency there is no net charge motion and foremost, it allows the study of particle dipole interactions. We also note that, as we work in a microgravity environment, there is no settling due to gravity in our experiments.\\

\begin{figure}
    \center
	\includegraphics[width=\columnwidth]{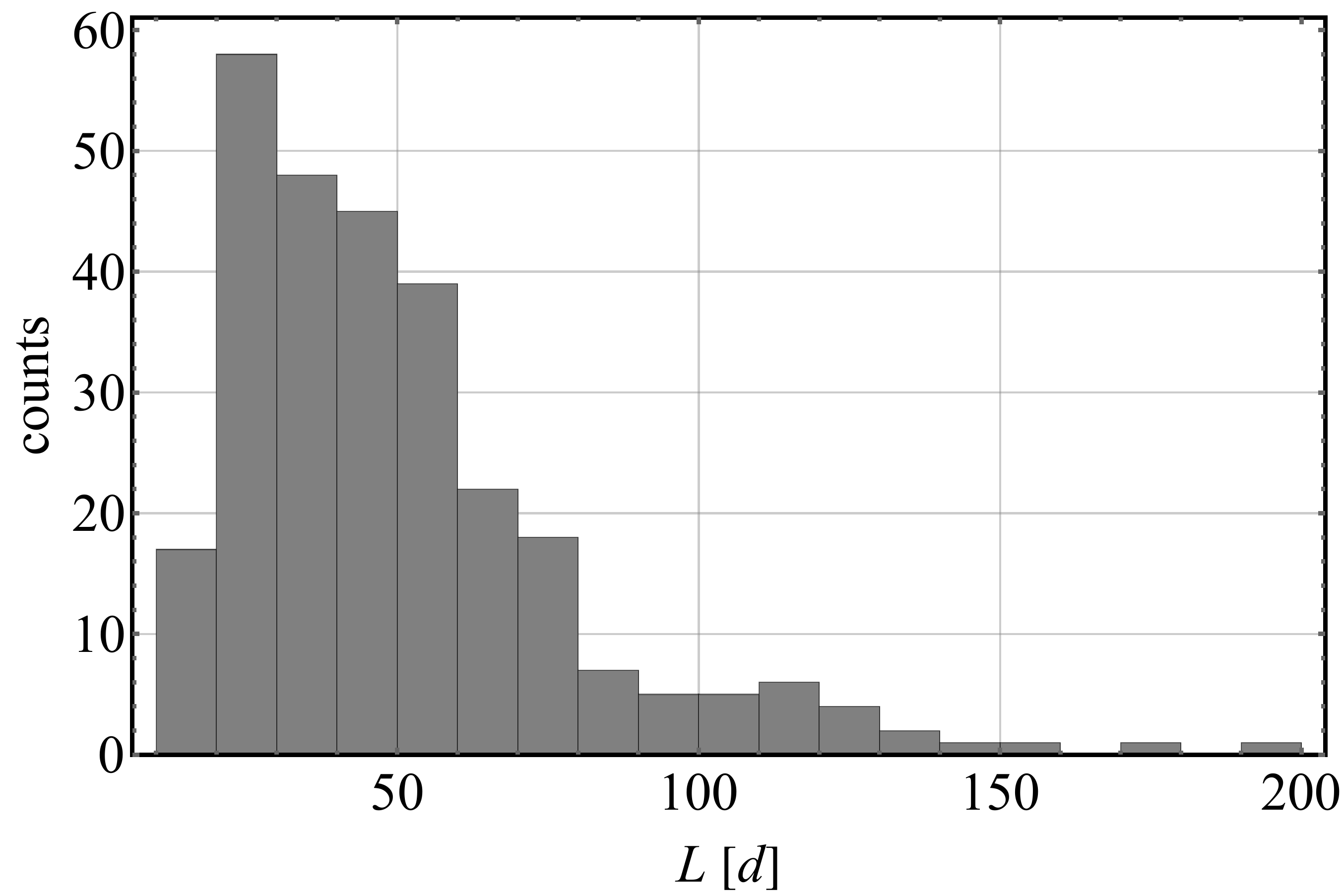}
	\caption{\label{fig.histo} Chain lengths $L$ in field direction in units of the particle diameter $d$ for the aggregates shown in Fig.~\ref{fig.chain-process} B.}
\end{figure}

We observed that, once particle chains have formed, they even persist when the field is switched off. Fig.~\ref{fig.histo} shows the distribution of the lengths of individual chains aligned with the field direction measured end-to-end. The extent of the chains, as derived by 2D image analysis, exceeds 100 particle diameters.

We note that the maximum size measured is not a general limit in the size of aggregates. It just mirrors the limited local particle supply. As soon as all individual particles are embedded into chains, growth gets stalled. 
Therefore, Fig.~\ref{fig.histo} does not allow a statistical analysis of the chain formation process.

\subsection{Aggregation model}
\label{chp.agg-model}

Induced dipole attraction is the most plausible explanation for the observation of particle chains. To make this more quantitative, a simple model should be capable of reproducing the observed collision timescale or aggregation timescale. 
We start with the number density $n$ of grains or a number of $N$ grains in a volume $V$. The particle launching results in a confined stream of particles due to the interaction between grains and gas. As a representative example, the initial particle density is determined in the white marked area in Fig.~\ref{fig.chain-process} (top). 
As the observation is limited to 2D, the local volume of the particle cloud is assumed to be cylindrical with 

\begin{equation}
V_{\rm{c}}=\pi\, \frac{d^2}{4}\,h. 
\end{equation}

Here, $d$ is the diameter and $h$ is the height of the cylindrical volume and their values are taken graphically from the images (Fig. \ref{fig.chain-process}). 
The number of particles $N$ in the marked area can be estimated using the image gray scale, as the mean gray scale of an image section scales with the particle number density within the section. To calibrate this dependence we chose image sections with smaller number densities, allowing to count all particles within the image section to determine the particle number density exactly. The background value is set to 255, while 0 is the value of a totally black pixel. This gives a dependence of the gray value of the image and the particle number. Since the particles are very transparent, the value can be scaled up to the more dense region. This results in a particle number density for the marked area $n=N/V_{\rm{c}}=2.5\,\rm{mm}^{-3}$, which gives an initial particle spacing of $x=1/\sqrt[3]{n}=0.74\,\rm{mm}$. The uncertainty of this average particle spacing is estimated as follows. Counting of the grains in the reference cell ($N_{\rm R}$) induces about 10\% uncertainty. The gray scale value of the reference cell ($g_{\rm R}$) is accurate to about  30\% due to fluctuating background. The diameter of the cylinder ($d$) is uncertain to about 10\% based on the reasonable maximum and minimum  values in view of missing information about the third dimension. In total we get an average spacing of $x=1/\sqrt[3]{n}=0.74 \pm 0.10\,\rm{mm}$.

With net charges on neighboring grains Coulomb attraction
would be
\begin{equation}
    F_{\rm{C}}=\frac{1}{4\pi\epsilon_0}\frac{q^2}{r^2}
\end{equation}
between two static charges $q$ of opposite polarity in a distance $r$. 

However, as chains only form after the field is turned on, grains are approaching each other due to a force tied to the external field
which then must be a field-induced dipole force. 
We only consider dipoles induced by the external field here as variations due to grain-grain interaction are secondary in comparison.
If two dipoles are aligned, the force on a dipole in the dipole-field of the other particle can be described by:

\begin{equation}
    F_{\text{dip}}= \dfrac{6}{4 \pi \epsilon_0} \, \dfrac{p^2}{r^4}
    \label{eq.dipol-force}
\end{equation}

Since the particles are of the same size and are exposed to the same electric field, both induced dipoles $p$ are the same.
The dipole moment of a dielectric material induced by a field $E$ is \citep{Jones1995}:
\begin{equation}
    p= 4 \pi \epsilon_0 K R^3 E
\end{equation}
Here, $R=82.5\,$ \textmu m and $\epsilon_0$ are the radius of the particles and the permittivity of vacuum. The Clausius-Mossotti function

\begin{equation}
    \label{eq.Clausius}
    K=\dfrac{\epsilon_{\rm{eff}} - 1}{\epsilon_{\rm{eff}} + 2}
\end{equation}

reflects the strength of the polarization of a spherical particle with an effective permittivity $\epsilon_{\rm{eff}}$. In the case of hollow spheres $\epsilon_{\rm{eff}}$ equals $0.155\,\epsilon_{\rm{r}}$ with $\epsilon_{\rm{r}}=7.3$ for soda-lime glass \citep{Jones1995}. We note that $K$ is larger by a factor 16 for compact particles. Additionally, the dielectric constant $\epsilon_r$ is not known exactly as the hollow spheres consist of a mixture of soda-lime glass and borosilicate glass, while the solid spheres used consist of soda-lime glass only.

\begin{figure}
	\includegraphics[width=\columnwidth]{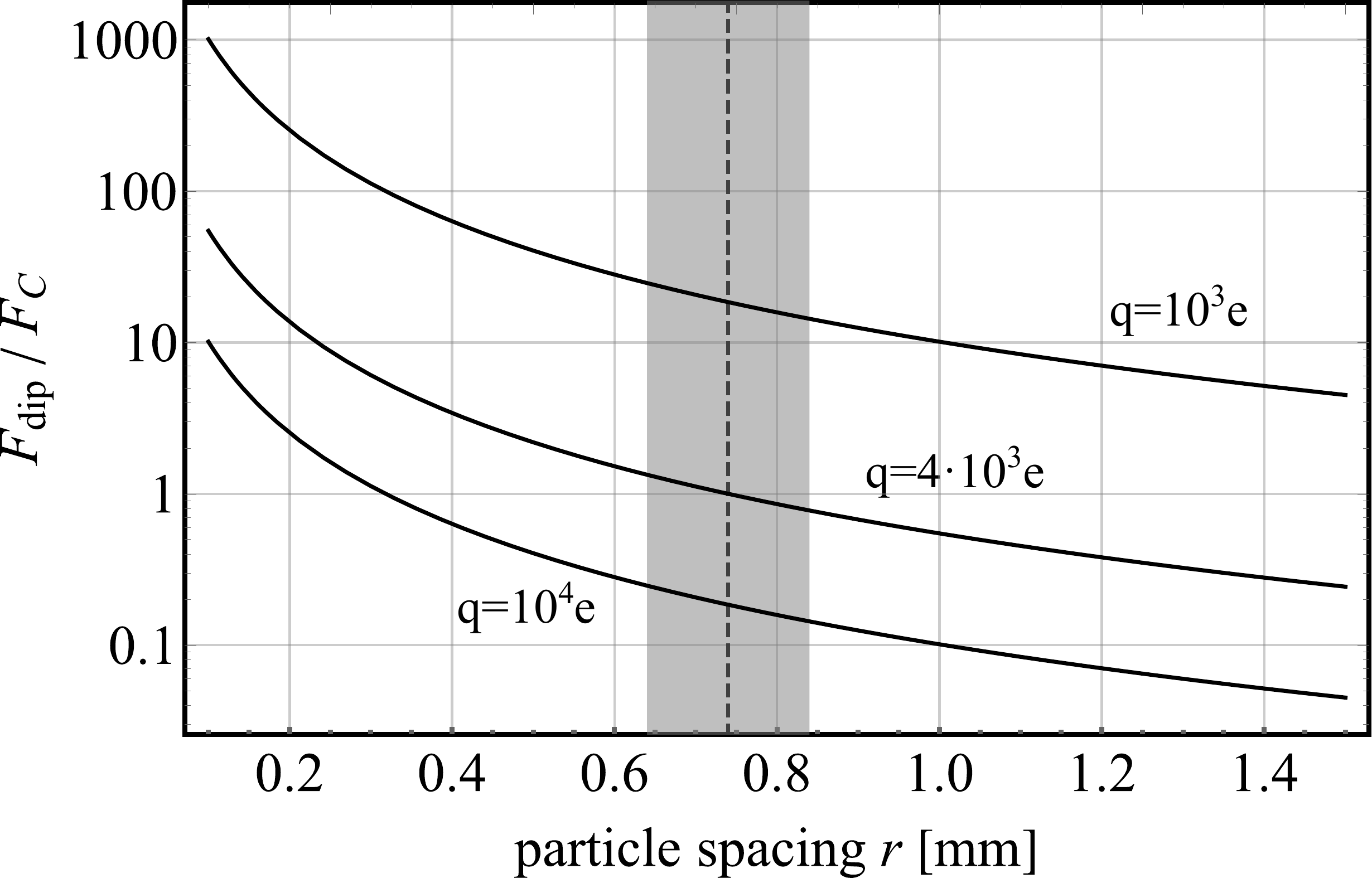}
	\caption{\label{fig.forces} Ratio of the force between induced dipoles and the Coulomb force for different net charges. The dashed line and the shaded area indicate the calculated distance $x=0.74\pm0.10\,\text{mm}$ in the particle cloud.}
\end{figure}

Fig.~\ref{fig.forces} shows the ratio of the dipole forces over net charge Coulomb forces for two single particles depending on distance and for two different net charges. Due to the $r^4$-dependence the induced dipoles get more dominant as the sample particles approach each other. At the average distance $x$ in the particle cloud the force is about $F_{\text{dip}}= 10^{-14}$\,N which is 20 times higher than the pure Coulomb force $F_{\text{C}}$ for a net charge of $q=10^3$\,e per particle. For a significantly larger charge in the range of $10^4$\,e the Coulomb interaction would be the driving force among the particles. As we already know that without electric field no interaction of the particles can be observed, the average net charge should be on the order of $10^3$\,e or below. This will be confirmed in section \ref{sec.net-charge}.

To estimate the timescale of collisions we integrated numerically the equation of motion, driven by the force of the induced dipoles, using the midpoint method. Two particles with distance $r$ approach each other and are accelerated by the dipole force $F_{\mathrm{dip}}$.

\begin{equation}
    m \cdot \ddot{r}(t) = F_{\mathrm{dip}}(r)
\end{equation}

\begin{figure}
    \center
	\includegraphics[width=\columnwidth]{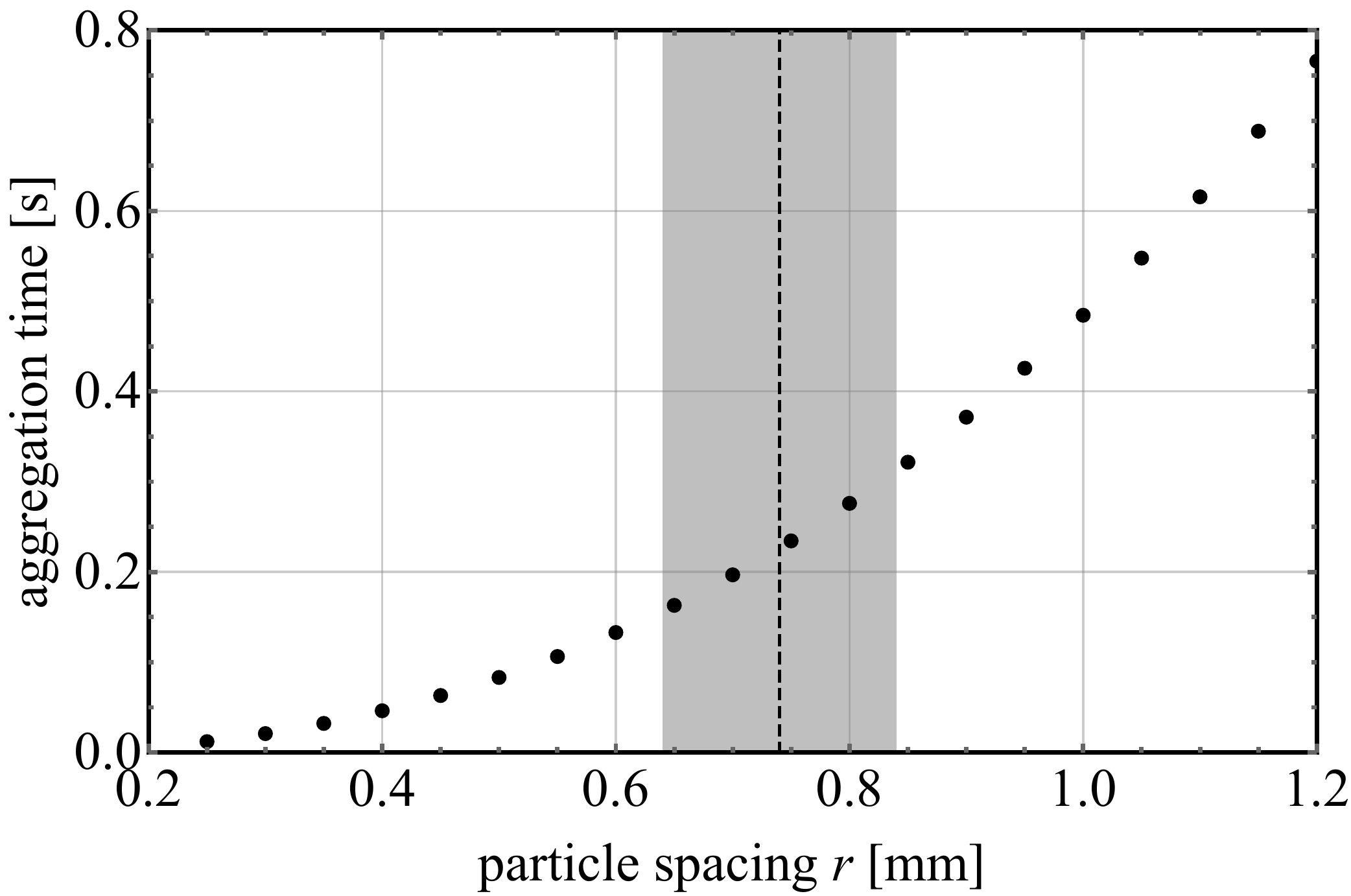}
	\caption{\label{fig.agg-time} Aggregation time (collision time for two spheres) depending on the initial spacing of the particles. The dashed line and the shaded area show the distance $x$ including its deviation in the example volume from Fig.~\ref{fig.chain-process}.}
\end{figure}

Here, $m$ is the mass of a particle, $\ddot{r}(t)$ the second order time derivative of the particle spacing and $F_{\mathrm{dip}}$ the dipole force.
Fig.~\ref{fig.agg-time} shows the collision time depending on the particle spacing. The average particle distance of $x=0.74$\,mm yields a collision time of 0.23\,s. Taking the lower limit of the uncertainty interval we get an aggregation time of 0.16\,s. In the experiments, we observe first collisions and chain formation in the highlighted area within 0.15\,s, which is in good agreement. 
Overall, the aggregation time seems to be independent of the position within the cloud which implies that the particle densities do not vary that much. Within its uncertainties, this simple model matches with the observed timescales, quantifying the influence of induced dipoles. 

\subsection{Measurement of net charges \label{sec.net-charge}}

\begin{figure}
	\includegraphics[width=\columnwidth]{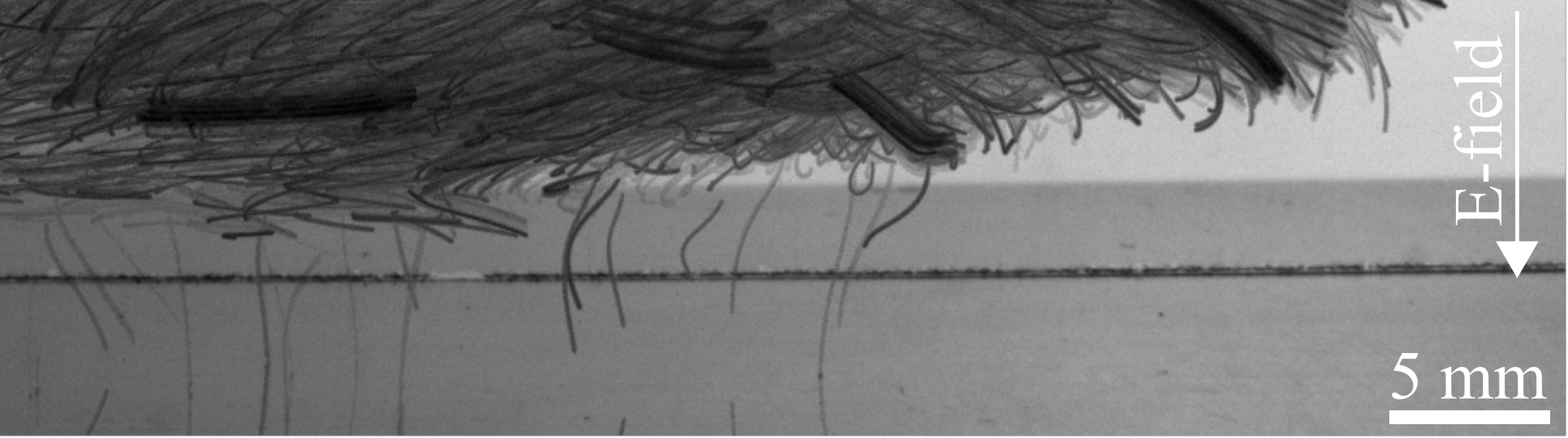}
	\caption{\label{fig.net-track} Overlay of images after turning on the electric field. 150 images are superimposed covering a time span of 1.5 s. While the majority of the spheres are moving in the cloud only a few spheres are accelerated towards the electrodes. Those are the heavily charged ones. The motion perpendicular to the electric field is the residual motion from the sample being injected into the capacitor.}
\end{figure}

To verify that net charges are small, we measured the net charge of single spheres as they move in a constant electric field in another experiment. The hollow spheres are entering the capacitor and after 3\,s of microgravity a constant field of 20\,kV/m is switched on. Although there are several thousand spheres in the volume, only a few show measurable motion towards the electrode. These spheres were tracked to determine their charge (Fig.~\ref{fig.net-track}). The gas-grain coupling time $\tau$ can be estimated using

\begin{equation}
    \tau=\dfrac{m}{6 \pi R \eta} \approx 6\, \text{ms},
    \label{eq.tau}
\end{equation}

which results from the equilibrium of inertia and Stokes drag \citep{Stokes1851}. Here, $m$ is the grain mass, $R$ the grain radius and $\eta=1.8\cdot 10^{-5}$\,Pa\,s the viscosity of air. This time is short compared to the total flight time until reaching the electrode. Therefore, the particles can be considered to move with terminal velocity $v$. For a single spherical grain with Stokes drag this is

\begin{equation}
v = \frac{q E}{m} \cdot \tau 
\label{eq.stokes}
\end{equation}

with the net charge $q$ and the electric field $E$. Combining both equations \eqref{eq.tau} and \eqref{eq.stokes} yields the net charge 

\begin{equation}
    q= \dfrac{6 \pi R \eta v}{E}.
    \label{eq.netcharge}
\end{equation}

Spheres carrying $10^3$\,e only reach a terminal velocity of 0.1\,mm/s which is slower than the movement of the whole particle cloud and would not be detected. We only identified 12 single spheres which carry charges high enough to be detected. Using equation~\eqref{eq.netcharge}, their charges are calculated to an average of $3 \cdot 10^4$\,e. In this range, Coulomb interaction is the dominant force at average distances (see Fig.~\ref{fig.forces}). However, the charge distribution of insulating glass spheres follows an exponential decay at the tails implying that there are a lot more little charged spheres than highly charged particles \citep{Haeberle2018, Steinpilz2020a}. In view of this and the non-detectable reaction of most of the cloud to the electric field, the measured values represent outliers of maximum charges in the sample while the majority of spheres is carrying $10^3$\,e or even less charge. Considering Fig.~\ref{fig.forces}, this confirms that the sample is dominated by field-induced dipoles.

Regarding charge densities, the grains reach values of $ 6\cdot 10^{-8}\,\textrm{C/m}^2$. Compared to glass and basaltic samples, this is one and two orders of magnitude lower, respectively \citep{Jungmann2018, Wurm2019, Steinpilz2020a}. As we compare different materials, this is no surprise, but additionally, it is unclear if the charging process works the same way for hollow spheres.
Anyway, we do not expect the dust particles to carry higher net charges than measured here due to their smaller size. Thus, dipoles would still be the driving force.

\section{Application to wind-driven matter}

The experiments were carried out with free particles in close proximity and show the importance of the attractive forces induced by external electric fields, especially in comparison to the case that particles are not attracted if no external field is present.

Larger grains have been studied in previous experiments where chains did also form in a non-alternating electric field (e.g. \citet{Jungmann2018, Steinpilz2020a}). This shows that the process also works with grains up to mm size and constant fields which are present in nature.
The field strength applied is on the order of field strengths measured in wind-driven dust and sand remobilization events on Earth \citep{Schmidt1998, Zheng2013}. As such, the experiments visualize that strong effects on aggregation might be expected after lofting events.
How does this ideal setting compare to conditions on other planetary surfaces? 

We first like to caution the reader that the dipole forces are still comparably small forces for grains on the ground. One may first deduce that induced dipoles change the force balance for particle lifting. This is not the case. Compared to cohesion or gravity, the dipole force is orders of magnitude smaller. We note that this does not say anything about the influence of net charges though, which is not accounted for in this work. 

Once aloft, the inertia of large sand-sized grains is still dominating over aligning and attractive effects of the induced dipoles. Due to large coupling times the relative velocities in a collision can be quite high and therefore, sand is unlikely to form chains.

However, the hollow spheres used in the experiments carried out here share low gas-grain coupling times with dust particles, that is they readily couple to the gas and only follow the gas flow afterwards. In nature, it depends on several factors like wind speed, grain shape or grain density, whether particles are lofted. But dust grains smaller than about 10\,\textmu m  go directly into suspension, leaving the saltation layer, eventually. If we scale the forces given above (Coulomb, Dipole, Clausius-Mossotti) for the hollow spheres to compact dust grains, dipole moments still dominate as particles approach each other. Therefore, once the dust particles are aloft it is likely that they  form chains instantly.

Dipole and Coulomb forces change with the particle size, that is their ratio scales with $R^2$ assuming a constant surface charge density. However, to compare the hollow glass beads used in the experiments with real dust samples considered as solid spheres, the dependence of the force ratio on $K^2$ has to be taken into account. For the total force ratio, this results in the same order of magnitude for the hollow glass spheres in the experiments and dust samples in the range of 10\,\textmu m. Therefore, we consider the hollow spheres as a suitable analog for real dust particles in terms of the importance of polarization over net charge. Additionally, the composition of the dust grains will also contribute to a difference between our model system and the real situation in nature. Different mineral phases (even within the grains) will influence the net charges of the grains as well as the charge distribution within the grains, leading to an extremely complex micro-physical system.

For strongly charged grains, \citet{Merrison2004} found evidence of electrostatic aggregation in laboratory experiments. In a less charged case as that studied here, dielectrophoresis by induced dipoles will also increase the collisional cross sections, enhancing the number of collisions. Dust grains will align with their long axis along the electric field and any new dust particle will be added preferentially on the long side producing chain-like aggregates. 

Alignment and chain formation only works as long as the external field is turned on. As cohesion is strong for dust particles, aggregates can keep this morphology, though they might evolve into different structures later on. Agglomerates higher up in an atmosphere or after sedimentation might break and therefore, they might not mirror the initial chain-like morphology anymore. 

We have to note that our model system is rather simple, considering slow gas-particles flows with monodispersed grain size distributions. It is not clear, to what degree chains are produced in a much more complex, polydisperse setting in a saltation layer of fast moving large grains. At least, collisional cross sections and cohesion forces are increased for dust particles. However, chain formation is not uncommon in directed electric and magnetic fields \citep{Shinbrot2006, Merrison2012, Kruss2018}. To some degree, chains will leave the saltation layer.

Our experiments are but one step in trying to understand the microphysics of dust particle lifting and atmospheric entrainment. The (chain-like) dust aggregation early after lifting might be significant in a wider context of atmospheric dust. 
At the low pressure of only a few mbar, lifting processes are way more challenging on Mars. Therefore, electrostatic effects may play a major role in particle remobilization, where wind or turbulence might suffice on Earth at first glance.
A detailed application of our findings to specific natural cases is beyond the scope of this paper but, again, an early chain-like structure can have a severe impact on the residence time of the dust in the atmosphere as well as on the radiative forcing, due to specific aerodynamical and optical properties of chain-like aggregates. 

\section{Conclusion}

In microgravity experiments, we observed the formation of chains made of hollow glass spheres by applying an external electric field. 
Using hollow glass spheres of 165\,\textmu m average diameter, chains reach a length of about 100 particles which makes up a length of a few cm. Without the application of an external electric field no aggregation is observed and therefore, we deduce that only induced dipoles can be the driving force for the formation of chain-like aggregates. 

We showed that the force caused by these dipoles in a dense region of a particle cloud can exceed the pure Coulomb forces by net charge attraction by far. Thus, net charges are estimated to a maximum of $10^4$ e and are not the most relevant property to describe the particle flow in an electric field in our case. This chain-like aggregation has not only been observed for hollow spheres but also for solid spheres with a size of nearly a mm, regardless of whether the field is constant or alternating. 

Our experiments are idealized and cannot be transferred one to one to natural settings. However, initial dust aggregates in natural settings with high electric fields might preferentially be elongated or chain-like. Even when the chains leave the electric field they keep their structure due to regular cohesion forces, which are related to van-der-Waals forces or more generally, surface energy, independent of the electric field. Once in contact, the cohesion forces need to be overcome to break the chains. Mostly, velocity gradients in a flow are small, therefore, gas drag acts on each grain in the chain the same way and shear forces will likely not destroy the chains.
If some chain structure is imprinted on dust aggregates, it will have significant influence on particle sedimentation due to the specific aerodynamical cross section, thus, dust may remain longer in the atmospheres. The presence of chains will influence the mass flux as well as the optical properties.

Quite generally, the microphysics of dust lifting and early aggregation will have impact on a wide scale from local dust storms on Earth to global storms on Mars. 
Any detail about the influence on the optical properties or sedimentation properties is beyond the scope of this paper though, as this requires a more complex model of lifting and aggregation. In view of the results of the experiments presented here, it will be worth to have a more closer look in future studies.

\section*{Acknowledgements}
This project is supported by DLR Space Administration with funds provided by the Federal Ministry for Economic Affairs and Energy (BMWi) under grant numbers DLR 50 WM 1762 and DLR 50 WM 2142.
This project has received funding from the European Union’s Horizon 2020 research and innovation program under grant agreement No 101004052. We also acknowledge the supporting works of Nils Völlings during the drop tower experiments and his preparative research on charged chains.\\

\def\aap{\textit{Astron. \& Astrophys.}}
\def\aapr{\textit{Astron. \& Asrophys. Rev.}}
\def\araa{\textit{Annu. Rev. Astro. Astrophys}}
\def\aaps{\textit{Astron. \& Asrophys. Suppl.}}
\def\adsr{\textit{Adv.~Space~Research}}
\def\aj{\textit{Astronomical~J.}}
\def\apj{\textit{Astrophysical~J.}}
\def\apjs{\textit{Astrophysical J. Suppl.}}
\def\apjl{\textit{Astrophysical J. Letters}}
\def\aspc{\textit{Astronomical Soc. Pacific Conf. Ser.}}
\def\ass{\textit{Astrophys. \& Space Sci.}}
\def\bala{\textit{Baltic Astronomy\/}}
\def\cmp{\textit{Contr. Mineralogy \& Petrology\/}}
\def\cpc{\textit{Computer Physics Comm.}}
\def\epsl{\textit{Earth \& Planetary Sci. Lett.}}
\def\gca{\textit{Geochimica \& Cosmochimica Acta\/}}
\def\ica{\textit{Icarus\/}}
\def\jp{\textit{J. Petrol.}}
\def\jqsrt{\textit{JQSRT\/}}
\def\mn{\textit{Monthly Notices Roy. Astr. Soc.\/}}
\def\mps{\textit{Meteoritics \& Planetary Sci.}}
\def\nat{\textit{Nature\/}}
\def\phre{\textit{Physical Rev.~E\/}}
\def\pre{\textit{Physical Rev.~E\/}}
\def\prl{\textit{Physical Rev. Lett.}}
\def\pss{\textit{Planetary \& Space Sci.}}
\def\rmf{\textit{Rev.Mod.Phys}}
\def\sci{\textit{Science\/}}
\def\ssr{\textit{Space Science Rev.}}
\def\mnras{\textit{Monthly Notices Roy. Astr. Soc.}}
\def\jgr{\textit{J. Geophys. Res.\/}}
\def\icarus{\textit{Icarus}}
\def\planss{\textit{Planetary and Space Science}}
\def\grl{\textit{Geophysical Research Letters}}
\def\apss{\textit{Astrophysics and Space Science}}


\bibliographystyle{apalike}
\bibliography{bibbi}

\begin{thebibliography}{}

\bibitem[Bagnold, 1941]{Bagnold1941}
Bagnold, R. (1941).
\newblock {\em The physics of blown sand and desert dunes}.
\newblock Meuthen.

\bibitem[{Bila} et~al., 2020]{Bila2020}
{Bila}, T., {Wurm}, G., {Onyeagusi}, F.~C., and {Teiser}, J. (2020).
\newblock {Lifting grains by the transient low pressure in a martian dust
  devil}.
\newblock {\em \icarus}, 339:113569.

\bibitem[Cimarelli et~al., 2014]{Cimarelli2014}
Cimarelli, C., Alatorre-Ibarg{\"u}engoitia, M., Kueppers, U., Scheu, B., and
  Dingwell, D.~B. (2014).
\newblock Experimental generation of volcanic lightning.
\newblock {\em Geology}, 42(1):79--82.

\bibitem[Esposito et~al., 2016]{Esposito2016}
Esposito, F., Molinaro, R., Popa, C.~I., Molfese, C., Cozzolino, F., Marty, L.,
  Taj-Eddine, K., Di~Achille, G., Franzese, G., Silvestro, S., and Ori, G.~G.
  (2016).
\newblock The role of the atmospheric electric field in the dust-lifting
  process.
\newblock {\em Geophysical Research Letters}, 43(10):5501--5508.

\bibitem[Feng, 2000]{Feng2000}
Feng, J.~Q. (2000).
\newblock Electrostatic interaction between two charged dielectric spheres in
  contact.
\newblock {\em Phys. Rev. E}, 62:2891--2897.

\bibitem[{Franzese} et~al., 2018]{Franzese2018}
{Franzese}, G., {Esposito}, F., {Lorenz}, R., {Silvestro}, S., {Popa}, C.~I.,
  {Molinaro}, R., {Cozzolino}, F., {Molfese}, C., {Marty}, L., and {Deniskina},
  N. (2018).
\newblock {Electric properties of dust devils}.
\newblock {\em Earth and Planetary Science Letters}, 493:71--81.

\bibitem[Greeley and Iversen, 1985]{Greeley1985}
Greeley, R. and Iversen, J.~D. (1985).
\newblock {\em Wind as a Geological Process on Earth, Mars, Venus and Titan}.
\newblock Cambridge University Press.

\bibitem[Haeberle et~al., 2018]{Haeberle2018}
Haeberle, J., Schella, A., Sperl, M., Schr{\"o}ter, M., and Born, P. (2018).
\newblock Double origin of stochastic granular tribocharging.
\newblock {\em Soft matter}, 14(24):4987--4995.

\bibitem[Harrison et~al., 2016]{Harrison2016}
Harrison, R.~G., Barth, E., Esposito, F., Merrison, J., Montmessin, F., Aplin,
  K.~L., Borlina, C., Berthelier, J.-J., D{\'e}prez, G., Farrell, W.~M., et~al.
  (2016).
\newblock Applications of electrified dust and dust devil electrodynamics to
  martian atmospheric electricity.
\newblock {\em Space Science Reviews}, 203(1-4):299--345.

\bibitem[James et~al., 2002]{James2002}
James, M.~R., Gilbert, J.~S., and Lane, S.~J. (2002).
\newblock Experimental investigation of volcanic particle aggregation in the
  absence of a liquid phase.
\newblock {\em Journal of Geophysical Research: Solid Earth}, 107(B9):ECV
  4--1--ECV 4--13.

\bibitem[Jones, 1995]{Jones1995}
Jones, T.~B. (1995).
\newblock {\em Electromechanics of Particles}.
\newblock Cambridge University Press.

\bibitem[Jungmann et~al., 2018]{Jungmann2018}
Jungmann, F., Steinpilz, T., Teiser, J., and Wurm, G. (2018).
\newblock Sticking and restitution in collisions of charged sub-mm dielectric
  grains.
\newblock {\em Journal of Physics Communications}.

\bibitem[Jungmann and Wurm, 2021]{Jungmann&Wurm2021}
Jungmann, F. and Wurm, G. (2021).
\newblock Observation of bottom-up formation for charged grain aggregates
  related to pre-planetary evolution beyond the bouncing barrier.
\newblock {\em A\&A}, 650:A77.

\bibitem[{Kahre} et~al., 2017]{Kahre2017}
{Kahre}, M.~A., {Murphy}, J.~R., {Newman}, C.~E., {Wilson}, R.~J., {Cantor},
  B.~A., {Lemmon}, M.~T., and {Wolff}, M.~J. (2017).
\newblock {\em {The Mars Dust Cycle}}, pages 229--294.

\bibitem[{Klose} and {Shao}, 2013]{Klose2013}
{Klose}, M. and {Shao}, Y. (2013).
\newblock {Large-eddy simulation of turbulent dust emission}.
\newblock {\em Aeolian Research}, 8:49--58.

\bibitem[{Klose} et~al., 2017]{Klose2017}
{Klose}, M., {Webb}, N., {Gill}, T.~E., {Van Pelt}, S., and {Okin}, G. (2017).
\newblock {Can dust emission mechanisms be determined from field measurements?}
\newblock In {\em EGU General Assembly Conference Abstracts}, EGU General
  Assembly Conference Abstracts, page 578.

\bibitem[Kok et~al., 2012]{Kok2012}
Kok, J.~F., Parteli, E. J.~R., Michaels, T.~I., and Karam, D.~B. (2012).
\newblock The physics of wind-blown sand and dust.
\newblock {\em Reports on Progress in Physics}, 75(10):106901.

\bibitem[Kok and Renno, 2008]{Kok2008}
Kok, J.~F. and Renno, N.~O. (2008).
\newblock Electrostatics in wind-blown sand.
\newblock {\em Phys. Rev. Lett.}, 100:014501.

\bibitem[Kruss et~al., 2020]{Kruss2020a}
Kruss, M., Musiolik, G., Demirci, T., Wurm, G., and Teiser, J. (2020).
\newblock Wind erosion on mars and other small terrestrial planets.
\newblock {\em Icarus}, 337:113438.

\bibitem[Kruss and Wurm, 2018]{Kruss2018}
Kruss, M. and Wurm, G. (2018).
\newblock Seeding the formation of mercurys: An iron-sensitive bouncing barrier
  in disk magnetic fields.
\newblock {\em The Astrophysical Journal}, 869(1):45.

\bibitem[Kruss and Wurm, 2020]{Kruss2020b}
Kruss, M. and Wurm, G. (2020).
\newblock Composition and size dependent sorting in preplanetary growth:
  Seeding the formation of mercury-like planets.
\newblock {\em The Planetary Science Journal}, 1(1):23.

\bibitem[Lacks and Sankaran, 2011]{Lacks2011}
Lacks, D.~J. and Sankaran, R.~M. (2011).
\newblock Contact electrification of insulating materials.
\newblock {\em Journal of Physics D: Applied Physics}, 44(45):453001.

\bibitem[{Lemmon} et~al., 2019]{Lemmon2019}
{Lemmon}, M.~T., {Guzewich}, S.~D., {McConnochie}, T., {de Vicente-Retortillo},
  A., {Mart{\'\i}nez}, G., {Smith}, M.~D., {Bell}, J.~F., {Wellington}, D., and
  {Jacob}, S. (2019).
\newblock {Large Dust Aerosol Sizes Seen During the 2018 Martian Global Dust
  Event by the Curiosity Rover}.
\newblock {\em \grl}, 46(16):9448--9456.

\bibitem[{Mahowald} et~al., 2014]{Mahowald2014}
{Mahowald}, N., {Albani}, S., {Kok}, J.~F., {Engelstaeder}, S., {Scanza}, R.,
  {Ward}, D.~S., and {Flanner}, M.~G. (2014).
\newblock {The size distribution of desert dust aerosols and its impact on the
  Earth system}.
\newblock {\em Aeolian Research}, 15:53--71.

\bibitem[Matias et~al., 2018]{Matias2018}
Matias, A. F.~V., Shinbrot, T., and Ara\'ujo, N. A.~M. (2018).
\newblock Mechanical equilibrium of aggregates of dielectric spheres.
\newblock {\em Phys. Rev. E}, 98:062903.

\bibitem[{M{\'e}ndez Harper} et~al., 2020]{MendezHarper2020}
{M{\'e}ndez Harper}, J., {Courtland}, L., {Dufek}, J., and {McAdams}, J.
  (2020).
\newblock {Microphysical Effects of Water Content and Temperature on the
  Triboelectrification of Volcanic Ash on Long Time Scales}.
\newblock {\em Journal of Geophysical Research (Atmospheres)}, 125(14):e31498.

\bibitem[{M{\'e}ndez Harper} et~al., 2021]{MendezHarper2021}
{M{\'e}ndez Harper}, J., {Dufek}, J., and {McDonald}, G.~D. (2021).
\newblock {Detection of spark discharges in an agitated Mars dust simulant
  isolated from foreign surfaces}.
\newblock {\em \icarus}, 357:114268.

\bibitem[Merrison et~al., 2008]{Merrison2008}
Merrison, J., Bechtold, H., Gunnlaugsson, H., Jensen, A., Kinch, K., Nornberg,
  P., and Rasmussen, K. (2008).
\newblock An environmental simulation wind tunnel for studying aeolian
  transport on mars.
\newblock {\em Planetary and Space Science}, 56(3):426 -- 437.

\bibitem[Merrison et~al., 2007]{Merrison2007}
Merrison, J., Gunnlaugsson, H., Nørnberg, P., Jensen, A., and Rasmussen, K.
  (2007).
\newblock Determination of the wind induced detachment threshold for granular
  material on mars using wind tunnel simulations.
\newblock {\em Icarus}, 191(2):568 -- 580.

\bibitem[{Merrison} et~al., 2004]{Merrison2004}
{Merrison}, J., {Jensen}, J., {Kinch}, K., {Mugford}, R., and {N{\o}rnberg}, P.
  (2004).
\newblock {The electrical properties of Mars analogue dust}.
\newblock {\em \planss}, 52(4):279--290.

\bibitem[{Merrison}, 2012]{Merrison2012}
{Merrison}, J.~P. (2012).
\newblock {Sand transport, erosion and granular electrification}.
\newblock {\em Aeolian Research}, 4:1--16.

\bibitem[Mueller et~al., 2017]{Mueller2017}
Mueller, S.~B., Kueppers, U., Ametsbichler, J., Cimarelli, C., Merrison, J.~P.,
  Poret, M., Wadsworth, F.~B., and Dingwell, D.~B. (2017).
\newblock Stability of volcanic ash aggregates and break-up processes.
\newblock {\em Scientific reports}, 7(1):1--11.

\bibitem[{Musiolik} et~al., 2017]{Musiolik2017}
{Musiolik}, G., {de Beule}, C., and {Wurm}, G. (2017).
\newblock {Analog Experiments on Tensile Strength of Dusty and Cometary
  Matter}.
\newblock {\em \icarus}, 296:110--116.

\bibitem[Musiolik et~al., 2018]{Musiolik2018}
Musiolik, G., Kruss, M., Demirci, T., Schrinski, B., Teiser, J., Daerden, F.,
  Smith, M.~D., Neary, L., and Wurm, G. (2018).
\newblock Saltation under martian gravity and its influence on the global dust
  distribution.
\newblock {\em Icarus}, 306:25 -- 31.

\bibitem[Rasmussen et~al., 2009]{Rasmussen2009}
Rasmussen, K.~R., Kok, J.~F., and Merrison, J.~P. (2009).
\newblock Enhancement in wind-driven sand transport by electric fields.
\newblock {\em Planetary and Space Science}, 57(7):804 -- 808.

\bibitem[Rasmussen et~al., 2015]{Rasmussen2015}
Rasmussen, K.~R., Valance, A., and Merrison, J. (2015).
\newblock Laboratory studies of aeolian sediment transport processes on
  planetary surfaces.
\newblock {\em Geomorphology}, 244:74 -- 94.
\newblock Laboratory Experiments in Geomorphology 46th Annual Binghamton
  Geomorphology Symposium 18-20 September 2015.

\bibitem[Renno et~al., 2004]{Renno2004}
Renno, N.~O., Abreu, V.~J., Koch, J., Smith, P.~H., Hartogensis, O.~K.,
  De~Bruin, H. A.~R., Burose, D., Delory, G.~T., Farrell, W.~M., Watts, C.~J.,
  Garatuza, J., Parker, M., and Carswell, A. (2004).
\newblock Matador 2002: A pilot field experiment on convective plumes and dust
  devils.
\newblock {\em Journal of Geophysical Research: Planets}, 109(E7).

\bibitem[{Renno} and {Kok}, 2008]{Renno2008}
{Renno}, N.~O. and {Kok}, J.~F. (2008).
\newblock {Electrical Activity and Dust Lifting on Earth, Mars, and Beyond}.
\newblock {\em \ssr}, 137(1-4):419--434.

\bibitem[Schmidt et~al., 1998]{Schmidt1998}
Schmidt, D.~S., Schmidt, R.~A., and Dent, J.~D. (1998).
\newblock Electrostatic force on saltating sand.
\newblock {\em Journal of Geophysical Research: Atmospheres},
  103(D8):8997--9001.

\bibitem[Shao and Lu, 2000]{Shao2000}
Shao, Y. and Lu, H. (2000).
\newblock A simple expression for wind erosion threshold friction velocity.
\newblock {\em Journal of Geophysical Research: Atmospheres},
  105(D17):22437--22443.

\bibitem[{Shinbrot} et~al., 2006]{Shinbrot2006}
{Shinbrot}, T., {Lamarche}, K., and {Glasser}, B.~J. (2006).
\newblock {Triboelectrification and Razorbacks: Geophysical Patterns Produced
  in Dry Grains}.
\newblock {\em \prl}, 96(17):178002.

\bibitem[{Steinpilz} et~al., 2020]{Steinpilz2020a}
{Steinpilz}, T., {Joeris}, K., {Jungmann}, F., {Wolf}, D., {Brendel}, L.,
  {Teiser}, J., {Shinbrot}, T., and {Wurm}, G. (2020).
\newblock {Electrical charging overcomes the bouncing barrier in planet
  formation}.
\newblock {\em Nature Physics}, 16(2):225--229.

\bibitem[Steinpilz et~al., 2020]{Steinpilz2020b}
Steinpilz, T., Jungmann, F., Joeris, K., Teiser, J., and Wurm, G. (2020).
\newblock Measurements of dipole moments and a q-patch model of collisionally
  charged grains.
\newblock {\em New Journal of Physics}, 22(9):093025.

\bibitem[Stokes, 1851]{Stokes1851}
Stokes, G.~G. (1851).
\newblock On the effect of the internal friction of fluids on the motion of
  pendulums.
\newblock {\em Transactions of the Cambridge Philosophical Society}, 9.

\bibitem[Swann et~al., 2020]{Swann2020}
Swann, C., Sherman, D.~J., and Ewing, R.~C. (2020).
\newblock Experimentally derived thresholds for windblown sand on mars.
\newblock {\em Geophysical Research Letters}, 47(3):e2019GL084484.
\newblock e2019GL084484 2019GL084484.

\bibitem[Telling and Dufek, 2012]{Telling2012}
Telling, J. and Dufek, J. (2012).
\newblock An experimental evaluation of ash aggregation in explosive volcanic
  eruptions.
\newblock {\em Journal of Volcanology and Geothermal Research}, 209-210:1 -- 8.

\bibitem[Telling et~al., 2013]{Telling2013}
Telling, J., Dufek, J., and Shaikh, A. (2013).
\newblock Ash aggregation in explosive volcanic eruptions.
\newblock {\em Geophysical Research Letters}, 40(10):2355--2360.

\bibitem[Von~Kampen et~al., 2006]{VonKampen2006}
Von~Kampen, P., Kaczmarczik, U., and Rath, H.~J. (2006).
\newblock The new drop tower catapult system.
\newblock {\em Acta Astronautica}, 59(1-5):278--283.

\bibitem[Waitukaitis et~al., 2014]{Waitukaitis2014}
Waitukaitis, S.~R., Lee, V., Pierson, J.~M., Forman, S.~L., and Jaeger, H.~M.
  (2014).
\newblock Size-dependent same-material tribocharging in insulating grains.
\newblock {\em Phys. Rev. Lett.}, 112:218001.

\bibitem[Wurm et~al., 2019]{Wurm2019}
Wurm, G., Schmidt, L., Steinpilz, T., Boden, L., and Teiser, J. (2019).
\newblock A challenge for martian lightning: Limits of collisional charging at
  low pressure.
\newblock {\em Icarus}, 331:103--109.

\bibitem[{Zhang} and {Zhou}, 2020]{Zhang2020}
{Zhang}, H. and {Zhou}, Y.-H. (2020).
\newblock {Reconstructing the electrical structure of dust storms from locally
  observed electric field data}.
\newblock {\em Nature Communications}, 11:5072.

\bibitem[Zheng, 2013]{Zheng2013}
Zheng, X.-J. (2013).
\newblock Electrification of wind-blown sand: Recent advances and key issues.
\newblock {\em The European Physical Journal E}, 36:138.

\end{thebibliography}

\end{document}